\documentclass[twocolumn, twoside,prl]{revtex4}

\pdfoutput=1

\usepackage{graphicx}

\newcommand{\ra}{\rangle}
\newcommand{\la}{\langle}
\newcommand{\beq}{\begin{equation}}
\newcommand{\eeq}{\end{equation}}
\newcommand{\barr}{\begin{eqnarray}}
\newcommand{\earr}{\end{eqnarray}}
\newcommand{\Ud}{U^{\dagger}}
\newcommand{\up}{\uparrow_x}
\newcommand{\dn}{\downarrow_x}
\begin{document}

\title{Multiple-time states and multiple-time measurements in quantum mechanics}
\author{Yakir Aharonov$^{1,2}$}
\author{Sandu Popescu$^{3,4}$}
\author{Jeff Tollaksen$^2$}
\author{Lev Vaidman$^1$}
\affiliation{$^1$ Raymond and Beverly Sackler School of Physics and Astronomy, Tel Aviv University,  %
Tel Aviv, Israel}
\affiliation{$^2$ Center for Quantum Studies, Department of Computational and Data Sciences and Department of Physics, George Mason University, %
4400 University Drive, Fairfax, Virginia 22030, USA}
\affiliation{$^3$ H.H.Wills Physics Laboratory, University of Bristol, %
 Tyndall Avenue, Bristol BS8 1TL, U.K.}
\affiliation{$^4$ Hewlett-Packard Laboratories, Stoke Gifford, %
 Bristol BS12 6QZ, U.K.}

\begin{abstract} We discuss experimental situations that consist of multiple preparation and measurement stages.
This leads us to a new approach to quantum mechanics. In particular, we introduce the idea of multi-time quantum
states which are the appropriate tools for describing these experimental situations. We also describe multi-time
measurements and discuss their relation to multi-time states. A consequence of our new formalism is to put
states and operators on an equal footing. Finally we discuss the implications of our new approach to quantum
mechanics for the problem of the flow of time. \end{abstract}

\date{}

\maketitle

\section{1. Introduction}

The main aim of this paper is to introduce a new type of quantum state, a ``multiple-time state".
We will also discuss multiple-time measurements and introduce the notion of ``multiple-time
measurement states" or ``multiple-time history states".

The simplest situation, namely two-time states (also called pre- and post-selected states) was first discussed
by Aharonov, Bergman and Lebowitz \cite{ABL} in 1964 and was extensively studied during the last two decades
\cite{pre_and_post}.  The idea of multi-time measurements and the first steps towards multi-time states were
discussed by Aharonov and Albert in \cite{multi-time-measurements}. The present paper is based on ideas
described in the (unpublished) PhD theses of Vaidman and Popescu \cite{phd}.

From a mathematical point of view, the ``state" of a physical system is nothing other than a compact description
of all the relevant information we have about that system. The usual quantum state is perfectly suited for the
simple situations studied routinely in quantum mechanics, namely experiments that consist of a preparation stage
followed by a measurement stage. The state $|\Psi\ra$ (or the density matrix $\rho$, if appropriate) contains
all the information. Based on it, we can predict the probabilities of any measurement. Of course, we may know
much more about the preparation stage than what is encoded in the state, such as details about the measuring
devices that were used or about the past history of the system, but as far as the measurement stage is concerned
everything is encapsulated in $|\Psi\ra$ (or $\rho$). It is in fact remarkable that for some systems only very
few parameters are needed, such as 3 real numbers for a spin 1/2 particle, while we might know many more things
about the preparation (such as the magnetic field that may have acted on the spin during its entire history).

In any case, while the usual quantum state is perfectly suitable for describing the standard experiment as
discussed above, we can imagine more complex experiments that consist of many stages of preparation inter-spread
with  many stages of measurement (fig. 1). Multiple-time states refer to these situations.

To avoid any confusions, we want to emphasize from the outset that we do not want to modify quantum
theory. Our results are totally and completely part of ordinary quantum mechanics. Furthermore, we
want to make it clear that the ordinary formalism of quantum mechanics is perfectly capable of
describing every experiment that we consider here, including experiments that consist of many
preparation-measurement stages. The issue however is to get a convenient, compact and illuminating
description; as we will show, multiple-time states are ideally suited tools for this purpose.

We can, of course, consider such complex experiments in classical physics as well. In that case
however the experiment can always be decomposed into many elementary experiments, each involving a
single preparation-measurement stage, and there is effectively nothing interesting to note. Quantum
mechanically however, the situation is far more interesting.

In the discussion above we referred to quantum states as being simply the mathematical tools for describing the
system. However, states also have an ontological dimension. This is a highly debated issue, which even for the
simple case of a standard state $|\Psi\ra$ is very controversial. Does the state have a ``reality" of its own,
or is it just a mathematical tool for making predictions? Does the state actually collapse, or is the collapse
simply our updating the mathematical description following the acquisition of new data (the results of new
measurement).  Is the state a physical entity (such as in Bohm's pilot wave model)? Discussing the ontological
status of multi-time states is bound to be even more controversial. It is not our intention to dwell too much on
this issue here. Our main focus is simply to find out what are the parameters that describe the system fully;
the structure that we uncovered  is here to stay regardless of its interpretation. We will comment however in
the conclusions on our world view in the light of the present results.

\begin{figure}[h]
\includegraphics[scale=0.5]{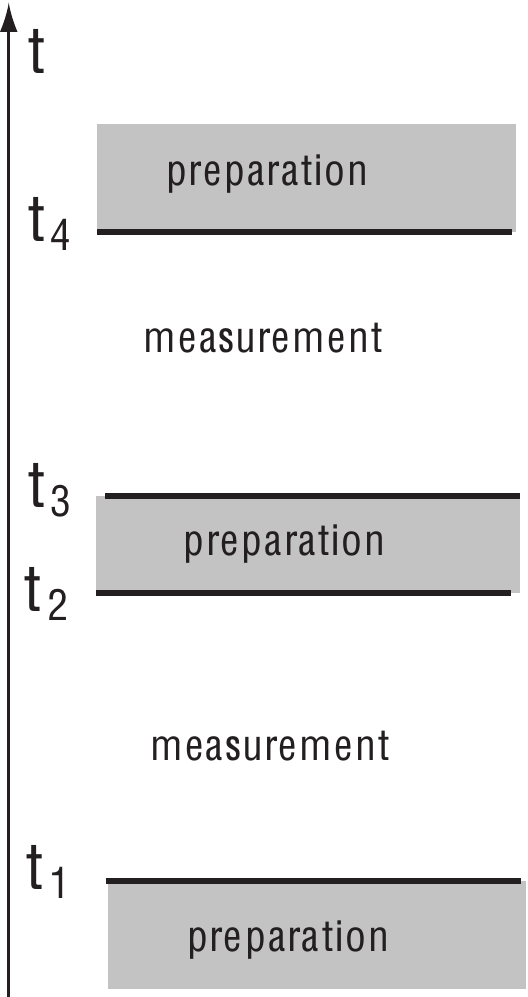} \caption {An experiment consisting of three "preparation" stages and two
"measurement" stages.}
\end{figure}

Coming now to measurements, we have two aims. The first is simply to discuss  ``multiple-time measurements".
These are measurements consisting of multiple measurement stages, but which cannot be decomposed into separate
measurements, one for each time. Considering such measurements is natural in the context of multi-time states.
Such measurements were introduced in \cite{multi-time-measurements}.  The second aim is to introduce the notion
of  ``measurement state" or, ``history state". Traditionally, the idea of ``state" is never associated with
measurements; it makes however a lot of sense. Indeed, consider first the notion of the ``state" of the system.
As discussed above, the state is nothing other than a compact description of all the relevant information about
a system,   the totality of the parameters needed to deduce what will happen to the system in any conceivable
situation. One may know much more about the system but this knowledge may be redundant. In a similar way, we can
ask what are all the relevant parameters that describe a measurement; the totality of these parameters will then
form a ``measurement state". For example, consider the usual ideal von Neumann measurement. Suppose we measure
an observable $A$. All the relevant information is encoded in the projectors $P_n$ on its eigen-subspaces. We
may know, of course, much more about the measurement (detailed information about the measuring device for
example) but this information is irrelevant. In fact, in theoretical discussions one very rarely discusses how
such a measurement could be performed - the explicit von Neumann measuring formalism is mostly restricted to a
few textbooks \cite{von_Neumann}. Then we can view the set of projectors $P_n$ as a ``state" describing the
measurement. We will call each individual projector a ``history state". While, of course, in this very simple
example the notion of history state is trivial, it's full force will become apparent when dealing with
multi-time measurements.

Again, one may ask  what is the ontological meaning of a ``measurement state". This is, of course, a perfectly
legitimate question. But whatever the ontological meaning is, from a formal point of view the set of projectors
$P_n$ are all that is needed to describe the ideal measurement, so they form a state. For the main part of this
paper we will focus on the mathematical formalism and discuss possible interpretations later.

\section{2. Simple two-time states: Pre-and post selection}

We will start with some simple situations and set up the general formalism afterwards. The physical situation,
illustrated in fig.2, is the following. There are two preparation stages, one at $t_1$ and the other at $t_2$,
and one measurement stage that takes place between $t_1$ and $t_2$. The system is prepared at time $t_1$ in some
quantum state $|\Psi\ra$. At a later time, $t_2$ the system is subjected to the measurement of an observable $B$
and the result $B=b$ is obtained; suppose that $b$ is a non-degenerate eigenvalue of $B$ corresponding to the
eigenstate $|\Phi\ra$. (i.e. $B|\Phi\ra=b|\Phi\ra$).

\begin{figure}[h]
\includegraphics[scale=0.5]{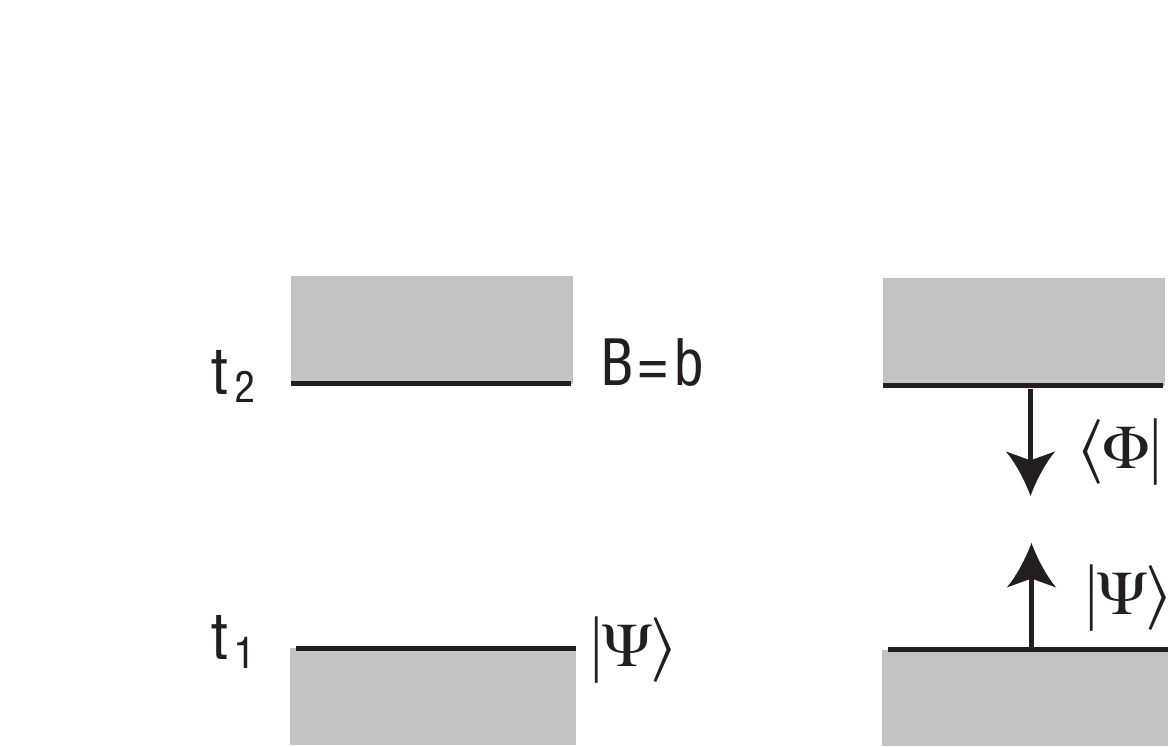} \caption {(a) At time $t_1$ the system is prepared in state $|\Psi\ra$ and at time $t_2$ an operator $B$ is measured. The outcome of $B$ happens to be $b$, the (non-degenerate) eigenvalue corresponding to the eigenstate $|\Phi\ra$.  (b) The same situation can be described by the two-time state $_{t_2}\la\Phi|~|\Psi\ra_{t_1}$}
\end{figure}

More precisely, there are in fact two interesting different physical situations that we can consider. In the
first case $t_1$ and $t_2$ are both in the past, $t_1$ in the remote past and $t_2$ in a more recent past. In
this case, the measurement of $B$ has already been performed and the result $b$ is the actual result of this
measurement. The other case  is when the second preparation did not yet take place. In this case we cannot
guarantee that the result $b$ will actually be obtained - the measurement might very well yield some other
result $b'$. Then, if we are interested only in the case in which the second preparation stage yields $b$, we
have no other option but discard the system  and start all over again. This is called ``post-selection".

In both the above cases, the entire physics for the period between the two preparation stages  is
governed by the two states $|\Psi\ra$ and $|\Phi\ra$. We now define a two-time state corresponding
to this situation by

\beq _{t_2}\la\Phi|~|\Psi\ra_{t_1}\label{preandpost} \eeq Note that the expression (\ref{preandpost}) is {\it
not} a scalar product (i.e. it is not the complex number $\la\Phi|\Psi\ra$ ) but a mathematical object which is
comprised of a bra and a ket vector with an empty slot in between. In this slot we eventually insert information
about the measurement period.

We use this state in the following way. Suppose that the particle evolves from $t_1$ to $t$
according to the unitary operator $U(t, t_1)$ and from $t$ to $t_2$ according to $U(t_2, t)$.
Furthermore, suppose that at $t$ the particle is subjected to an ideal, von Neumann measurement of
an observable $C$. Let $P_n$ be the projector associated to the eigenvalue $c_n$.

To obtain the probability ${\rm Prob}(C=c_n)$ that the measurement of the observable $C$ yields $C=c_n$ given
the two-time state (\ref{preandpost}) we simply insert the ``history"  $U(t_2,t)P_nU(t,t_1)$ in the available
slot, and make the contractions (i.e. we apply the operator $U(t,t_1)$ to $|\Psi\ra_{t_1}$, then act with the
projector $P_n$ etc.) to obtain the complex number \beq\la\Phi|U(t_2,t)P_nU(t,t_1)|\Psi\ra.\eeq The probability
${\rm Prob}(C=c_n)$ is then given by \beq {\rm Prob}(C=c_n)={1\over N}
\left|\la\Phi|U(t_2,t)P_nU(t,t_1)|\Psi\ra\right|^2\label{preandpostprob}.\eeq The normalization constant $N$ is
given by \beq N=\sum_k\left|\la\Phi|U(t_2,t)P_kU(t,t_1)|\Psi\ra\right|^2\eeq and ensures that the probabilities
for all possible outcomes add up to one. Note that the normalization constant $N$ could not have been included
in the definition of the state itself because its value depends not only on the state but also on the experiment
to which the state is subjected.

The case of multiple measurements performed between $t_1$ and $t_2$ can also be dealt with easily\cite{ABL}.
Consider for example two ideal von Neuman measurements of the observables $C$ and $D$ performed at $t$ and $t'$
respectively and let $P_n$ be the projector corresponding to $C=c_n$ and $Q_k$ be the projector associated to
$D=d_k$. Then

\barr &&{\rm Prob}(C=c_n \& D=d_k)=\nonumber\\&{1\over
N}&\left|\la\Phi|U(t_2,t')Q_kU(t',t)P_nU(t,t_1)|\Psi\ra\right|^2\label{multiple_measurements}\earr with

\beq N=\sum_{j,l}\left|\la\Phi|U(t_2,t')Q_jU(t',t)P_lU(t,t_1)|\Psi\ra\right|^2\eeq

\bigskip
\noindent

Finally, going beyond ideal von Neuman measurements, general measurements can be described in the POVM
formalism. Any measurement can be viewed as an interaction between the measured system and the measuring device,
followed by ``reading" the outcome indicated by the measuring device, i.e. by performing a von Neuman
measurement on the measuring device itself.  We will first discuss POVMs in the usual context of a one-time
state.

Consider first a ``detailed"  POVM. In such a measurement we leave no information unread. That is, we subject
the measuring device to a {\it complete} von Neuman  measurement, i.e. a von Neuman measurement which is such
that all the eigenvalues correspond to one-dimensional projectors. Following such a measurement, the system ends
up in a {\it pure} state - it may first get entangled with the measuring device but then the entanglement is
destroyed by reading the measuring device. (Note that as discussed below, following a general POVM, the system
may remain entangled with the measuring device.)

A detailed POVM is described by the operators that describe the evolution of a quantum state due to
the measurement. As noted above, under a detailed POVM pure states evolve into pure states. Let
$A_k$ be the operator that describes the evolution given the measurement outcome $k$, i.e. the
initial state $|\Psi\ra$ evolves into the (unnormalised) state $A_k|\Psi\ra$. \beq |\Psi\ra
\rightarrow A_k|\Psi\ra\eeq The operators $A_k$ are called Krauss operators. They are  linear
operators and they are arbitrary (not necessarily hermitian), up to the normalization condition
\beq \sum_k A_k^{\dagger}A_k=I\label{povm_normalization}\eeq where $I$ is the identity.  The probability of obtaining the
result $k$ is given by the norm of the post-measurement state, namely \beq
{\rm Prob}(k)=\la\Psi|A_k^{\dagger}A_k|\Psi\ra,\eeq and the normalization condition ensures that the
probabilities add up to 1.

Note that ideal von Neuman measurements are particular cases of detailed POVM's in which the Krauss operators
$A_k=P_k$ are projection operators. Time evolutions can also be easily included into the Krauss operators:
$A_k=U(t_2,t)P_kU(t,t_1)$ describes a von Neumann measurement preceded and followed by unitary time evolutions.
Furthermore, a series of ideal von Neuman measurements is also a particular detailed POVM. Indeed the operator
$U(t_2,t')Q_kU(t',t)P_nU(t,t_1)$ considered in (\ref{multiple_measurements}) above can be viewed as a Krauss
operator $A_{nk}$ corresponding to the outcome given by the pair ($k$,$n$). For simplicity,
for now on, unless explicitly specified otherwise, we consider the Krauss operators to cover the entire measurement
period they refer to.

Dealing with detailed POVM's in the context of pre-and-post selected states is identical to the way in which we
dealt with ideal von Neuman measurements. We associate Krauss operators $A_k$ with the entire experiment that
takes place between $t_1$ and $t_2$ (considering all unitary evolutions as part of the measurement itself) and
the probability of obtaining the result $k$ is given by \beq {\rm Prob}(k)={1\over N}
\left|\la\Phi|A_k|\Psi\ra\right|^2\label{krausspreandpostprob},\eeq where $N$ is a normalisation factor which
ensures that $\sum_k {\rm Prob}(k)=1$.

A general POVM is different from a detailed POVM in that we do not perform a {\it complete} reading of the
measuring device. To find the probabilities in this case, we can imagine that after finishing the original POVM
we proceed to read the remaining information but this new information is simply disregarded. (Since the
measuring device no longer interacts with the system, whether or not we make this supplementary reading of the
measuring device makes no difference to the system.) In effect, what we now have is a detailed POVM which is
such that to each outcome $k$ of the original POVM correspond a number of different outcomes $(k,\mu)$. All we
have to do, then, is simply add the probabilities for the different outcomes of the detailed POVM corresponding
to the same $k$.
Formally, to each measurement outcome $k$ of the original POVM correspond, in general, more
Krauss operators $A_{k\mu}$ where the index $k$ refers to the measurement outcome, and the index $\mu$ describes
different results that could have been differentiated but are lumped together and associated to the overall
outcome $k$. Again, the Krauss operators are arbitrary linear operators subject to the condition
$\sum_{k\mu}A_{k\mu}^{\dagger}A_{k\mu}=1$. For this POVM, the probability of obtaining the result $k$ is given
by \beq {\rm Prob}(k)={1\over N}\sum_{\mu}
\left|\la\Phi|A_{k\mu}|\Psi\ra\right|^2\label{generalkrausspreandpostprob},\eeq where $N$ is a normalisation
factor that ensures that $\sum_k {\rm Prob}(k)=1$.

We thus conclude that {\it any} measurements performed on a pre- and post selected system can be
described using the mathematical object $\la\Phi| |\Psi\ra$. We therefore are entitled to view
$\la\Phi| |\Psi\ra$ as {\it the} state of the system.

Up to this point, however, the situation is rather trivial and can be handled quite simply with the standard
formalism of quantum mechanics (in the manner indicated below). It suffices to consider the simplest case of a
single ideal von Neuman measurement discussed above; all other cases can be dealt with in a similar manner. For
simplicity, we will write $U_1=U(t,t_1)$ and $U_2=U(t_2,t)$. In the usual formalism we say that the system
starts in the state $|\Psi\ra$ and evolves into $U_1|\Psi\ra$ just prior to the measurement of $C$. The
probability to obtain $c_n$ is \beq \la\Psi|\Ud_1 P_nU_1|\Psi\ra\eeq and the state after the measurement becomes

\beq {{P_nU_1|\Psi\ra}\over{\sqrt{\la\Psi|\Ud_1 P_nU_1|\Psi\ra}}}\label{afterfirstmeasurement}.\eeq The
probability to obtain $b$, the eigenvalue corresponding to $|\Phi\ra$ when measuring $B$ at $t_2$ is the
absolute value square of the scalar product between $|\Phi\ra$ and the state (\ref{afterfirstmeasurement}) after
it undergoes propagation by $U_2$, i.e. \beq \left|{{\la\Phi|U_2P_nU_1|\Psi\ra}\over{\sqrt{\la\Psi|\Ud_1
P_nU_1|\Psi\ra}}}\right|^2\eeq

The overall probability to obtain $c_n$ and then $b$ is \barr&&
\left|{{\la\Phi|U_2P_nU_1|\Psi\ra}\over{\sqrt{\la\Psi|\Ud_1 P_nU_1|\Psi\ra}}}\right|^2 \la\Psi|\Ud_1
P_nU_1|\Psi\ra =\nonumber\\&=&\left|\la\Phi|U_2P_nU_1|\Psi\ra\right|^2.\earr This, however, is the probability
to obtain $c_n$ and then $b$. The {\it conditional} probability to obtain $c_n$ {\it given} that the measurement
of $B$ obtained $b$ is given by the usual conditional probability formula, by dividing the above probability by
the overall probability to obtain $b$ (given that we measured $C$ at $t$), i.e. \beq
{{\left|\la\Phi|U_2P_nU_1|\Psi\ra\right|^2}\over{\sum_k\left|\la\Phi|U_2P_kU_1|\Psi\ra\right|^2}}\eeq which
yields our formula (\ref{preandpostprob}). The case of multiple measurements can be handled in a similar way.

Note however that in this standard way of computing we use the notion of ``state" in an ontological way, not as
a repository of all the relevant information about the system. That is, we considered that the system actually
{\it is} in a state $|\Psi\ra$ at time $t_1$, that the state evolves into $U_1|\Psi\ra$ just prior to the
measurement, that it then collapses into $P_nU_1|\Psi\ra$ and so on. Of course, this usage may seem very
appealing from the point of view of an intuition established in standard discussions about quantum experiments.
Nevertheless, conceptually this is a very different usage of the notion of state.
In the situation in which we
are interested, when we have information about the system at two different times, the two-time state
(\ref{preandpost}) is the {\it only} mathematical object that can be called a state in the sense of containing
all the relevant information. The full power of this approach will become evident in the next section.

\section{3. Two-time states}

Let us now return to the two-time state (\ref{preandpost}). Although the case of pre-and post selection
described above can be dealt with relatively simply by the ordinary formalism, our formalism which uses the
two-time state $ _{t_2}\la\Phi|~|\Psi\ra_{t_1}$ has advantages. Not only  is it  more compact, but it also leads
us to ask new questions that could not be easily articulated in the old language.

The two-time state is a mathematical object living in a Hilbert space ${\cal H}={\overrightarrow{\cal
H}}_{t_2}\otimes{\overleftarrow{\cal H}}_{t_1}$ where ${\overleftarrow{\cal H}}_{t_1}$ is the Hilbert space of
the states at $t_1$ and ${\overrightarrow{\cal H}}_{t_2}$ is the Hilbert space for $t_2$. The arrows indicate
that ${\overleftarrow{\cal H}}_{t_1}$ is a space of ket vectors while ${\overrightarrow{\cal H}}_{t_2}$ is a
space of bra vectors.

The remarkable thing about two-time states is that, similar to ordinary quantum states, we can form
superpositions \cite{generalized}. In other words, any vector in  ${\cal H}={\overrightarrow{\cal
H}}_{t_2}\otimes{\overleftarrow{\cal H}}_{t_1}$ is a possible state of the system.

Consider the state

\beq \alpha_1{~}_{t_2}\la\Phi_1|~|\Psi_1\ra_{t_1} +\alpha_2{~}_{t_2}\la\Phi_2|~|\Psi_2\ra_{t_1}
\label{generalized_two_time}\eeq where $\la\Phi_1|$, $\la\Phi_2|$, $|\Psi_1\ra$ and $|\Psi_2\ra$ are arbitrary
states. What does this state represent and how can we prepare it?  The answer to this question is obtained by
looking at the probabilities for different measurements when the system is in this state.  Suppose that at time
$t$, between $t_1$ and $t_2$ we measure an observable $C$ and let the projection operator corresponding to the
eigenvalue $c_n$ be denoted by $P_n$.  Applying the rule used for simple two-time states, the probability for
obtaining $c_n$ is given by

\barr &&{\rm Prob}(C=c_n)= \nonumber\\&{1\over N}& {\left|
\alpha_1\la\Phi_1|U_2P_nU_1|\Psi_1\ra+\alpha_2\la\Phi_2|U_2P_nU_1|\Psi_2\ra\right|^2} \earr One way to prepare a
two-time state that leads to this result is the following. Consider our system and a supplementary particle, an
ancilla.  Consider now an ordinary pre- and post-selection as described before, but this time let both the
pre-selected state and the post-selected state be entangled states between the system and the ancilla.
Specifically, let the state at $t_1$ be \beq |\Psi_1\ra^S_{t_1}|1\ra^A_{t_1}+|\Psi_2\ra^S_{t_1}|2\ra^A_{t_1}\eeq
and the state at $t_2$ be \beq\alpha_1~^A_{t_2}\la1| ^S_{t_2}\la\Phi_1|+\alpha_2~ ^A_{t_2}\la2|
^S_{t_2}\la\Phi_2|,\eeq where the indices $S$ and $A$ denote the system and the ancilla. The two-time state for
the two particles is then \beq \left(\alpha_1~^A_{t_2}\la1| ^A_{t_2}\la\Phi_1|+\alpha_2~
^A_{t_2}\la2|(^S_{t_2}\la\Phi_2|\right)~~\left(|\Psi_1\ra^S_{t_1}|1\ra^A_{t_1}+|\Psi_2\ra^S_{t_1}|2\ra^A_{t_1}\right)\eeq
Suppose now that we perform a measurement on the system while the ancilla is left completely undisturbed - no
measurement is performed on it, and its Hamiltonian is zero. Since neither the projection operator $P_n$
associated with the measurement that is performed on the system nor the unitary evolutions $U_1$ and $U_2$
affect the ancilla, we obtain \barr && {\rm Prob}(C=c_n)=\nonumber\\&{1\over N}&\left|\left(\alpha_1\la1|
\la\Phi_1|+\alpha_2\la2|\la\Phi_2|\right)U_2P_nU_1\left(|\Psi_1\ra|1\ra+|\Psi_2\ra|2\ra\right)\right|^2\nonumber\\&=&{1\over
N}{\left|\alpha_1\la\Phi_1|U_2P_nU_1|\Psi_1\ra+\alpha_2\la\Phi_2|U_2P_nU_1|\Psi_2\ra\right|^2} \earr

So as long as we are interested in the system alone and trace over the ancilla the system is
described by (\ref{generalized_two_time}). The state (\ref{generalized_two_time}) is a {\it pure,
entangled} two-time state. The entanglement is between the states of the system at the two
different moments of time, more precisely between the ``forward in time" propagating states prepared
at $t_1$ and the ``backward in time" propagating states prepared at $t_2$.

Note that there are many other - in fact infinitely many other ways - in which the state
(\ref{generalized_two_time}) can be prepared. For example we can pre-select

\beq \sum_{i=1,2} \beta_i |\Psi_i\ra^S_{t_1}|i\ra^A_{t_1}\label{gen_two_time_prep_one}\eeq and post-select \beq
\sum_{j=1,2} \gamma_j {}_{t_2}^S \la\Phi_i|{}_{t_2}^A\la j|\label{gen_two_time_prep_two}\eeq with
$\gamma_i\beta_i=\alpha_i$. This freedom in preparing the state (\ref{generalized_two_time}) is similar to
freedom in the way in which an ordinary density matrix for a system can be obtained by entanglement with an
ancilla - there are infinite many pure entangled states that lead to the same reduced density matrix for the
system.

The generalization of state (\ref{generalized_two_time}) and of its method of preparation
(\ref{gen_two_time_prep_one}), (\ref{gen_two_time_prep_two}) to a superposition with an arbitrary
number of terms is obvious.

Yet another way to prepare arbitrary superpositions of two-time states is to put all information about the
two-time state in the initial state of the system and ancilla and to use a standard post-selection to transfer
information from the ancilla onto the system (fig.3). The simplest way to describe this method is to use a
decomposition of the desired two-time state using orthonormal basis vectors $_{t_2}\la j| |i\ra_{t_1}$ in ${\cal
H}={\overrightarrow{\cal H}}_{t_2}\otimes{\overleftarrow{\cal H}}_{t_1}$. Consider an arbitrary two-time state
\beq\sum_{i,j}\alpha_{ij}~ {}_{t_2}\la j| |i\ra_{t_1}.\label{general_basis-prepost} \eeq To prepare this state
we start with our quantum system and an ancilla in the pre-selected state \beq \sum_{i,j}\alpha_{ij}
|i\ra^S_{t_1} | j\ra^A_{t_1}\eeq which is a  ``map" of the desired state. We then post-select the maximally
entangled state \beq {}^{SA}_{t_2}\la\Phi^+|=\sum_n {}^S_{t_2}\la n|~{}^A_{t_2}\la
n|.\label{maximally_entangled}\eeq This can be done, for example, by measuring the well known Bell operator and
selecting the appropriate result. By post-selecting the maximally entangled state $|\Phi^+\ra$, we effectively
transfer the state of the ancilla into the backward-in-time propagating state of the system. In other words,
post-selecting on the maximally entangled state $|\Phi^+\ra$ acts as a channel  by which a ket vector of the
ancilla is transformed into a bra vector of the system (see fig 3).
\begin{figure}[h]
\includegraphics[scale=0.8]{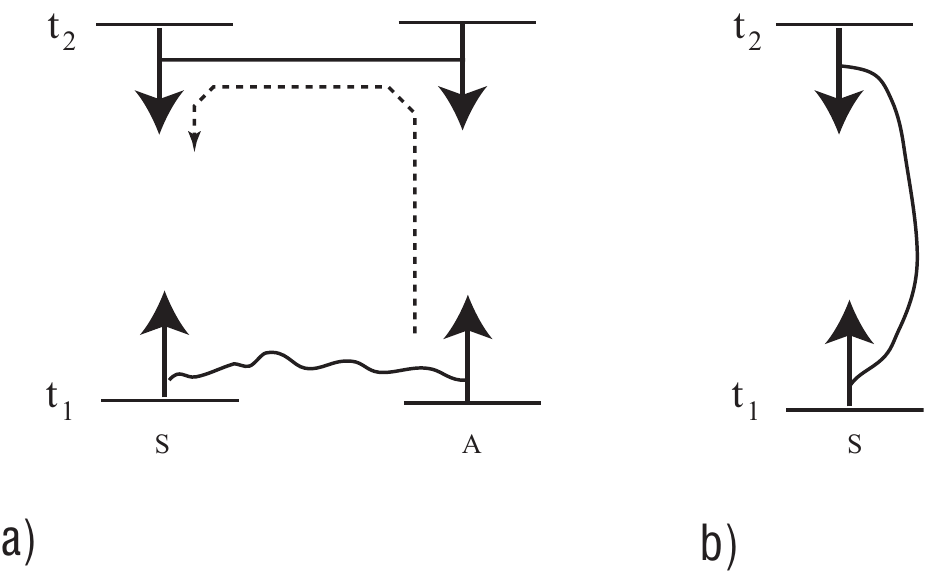} \caption {(a) The system S interacts with the ancilla A. The arrows represent
states propagating "forward" and "backward" in time,i.e ket and bra vectors. The wiggled line connecting the
forward in time propagating states (i.e. ket vectors) describes (arbitrary) entanglement. The continuous line
connecting the backward in time propagating states (i.e. bra vectors) illustrates maximal entanglement. More
precisely, since it refers to the bra vectors, it denotes post-selecting the maximally entangled state for the
system and the ancilla. The dotted line illustrates how entanglement is transferred from the ancilla onto the
system. The diagram (b) illustrates the same situation as(a)but from the point of view of the system alone. }
\end{figure}
Indeed, the pre-and post selected state of system+ancilla is \beq \sum_{ij}\alpha_{ij} {} ^{SA}\la
\Phi^+|~|i\ra^S |j\ra^A.\label{general_basis-prepost_ancilla}\eeq  When only measurements on the system are
concerned, we can contract the ancilla states obtaining \barr \sum_{ij}\alpha_{ij} {} ^{SA}\la \Phi^+|~|i\ra^S
|j\ra^A&=&\sum_{ijn}\alpha_{ij} {} ^A\la n|{}^S\la n| ~|i\ra^S |j\ra^A=\nonumber\\ \sum_{ijn}\alpha_{ij} {}^S\la
n| ~|i\ra^S {} ^A\la n|j\ra^A&=&\sum_{ijn}\alpha_{ij} {}^S\la n| ~|i\ra^S \delta_{nj}=\nonumber\\
\sum_{i,j}\alpha_{ij}~{}_{t_2}\la j| |i\ra_{t_1},\earr which is the desired state (\ref{general_basis-prepost}).

Until now, we discussed  two-time states of  a single quantum system. Of course, any number of particles can be
grouped together into a single system, so the discussion was completely general. We may however find it
convenient to describe different particles separately.  Consider for example a quantum system composed of two
particles, $A$ and $B$. A general pure two-time state is

\beq \sum_{ijkl}\alpha_{ijkl} ~_{t_2}^A\la i|_{t_2}^B\la j|~|k\ra^A_{t_1}|l\ra^B_{t_1}\eeq In general, such a
state is entangled both between the two particles, as well as between the two times. For example there are
states in which the post-selected state of particle A is entangled with the preselected state of particle B,
etc.

Finally, we note that along with pure two-time states we can have {\it mixed} two-time states. A
mixture arises when we prepare different pure two-time states with different probabilities.

\section{4. Multiple-time states}

The two-time states discussed above are just  the simplest example of multiple-time states. They
correspond to the situation in which there is one measurement stage sandwiched between two
preparation stages, as illustrated in fig 1.  Our formalism however applies equally well to
situations consisting of multiple preparation and measurement stages.

Consider an experiment as illustrated in fig. 4. To each time boundary between a preparation period followed
by a measurement period we associate a Hilbert space of ket vectors and to each time boundary between a
measurement period followed by a preparation period we associate a Hilbert space of bra vectors. The total
Hilbert space is the tensor product of the Hilbert spaces for all the time boundaries,  ${\cal H}={\cal
H}_{t_n}\otimes\dots\otimes{\overrightarrow{\cal H}}_{t_{k+1}}\otimes{\overleftarrow{\cal
H}}_{t_k}\otimes{\overrightarrow{\cal H}}_{t_{k-1}}\otimes\dots\otimes{\cal H}_{t_1}$. Note that the bra and ket
Hilbert spaces alternate due to the alternation of preparation and measurement periods. Furthermore, note also
that we did not explicitly mark arrows on the first and last Hilbert space. This is because there are four
different cases (fig 4) depending on whether the first and last Hilbert spaces are bra or ket spaces i.e.
whether the procedure starts (ends) with a preparation or measurement period. Which of these four cases occurs
depends on whether the past and future are uncertain or well-defined.  We will discuss the significance of the
difference between these four cases shortly.

\begin{figure}[h]
\includegraphics[scale=0.5]{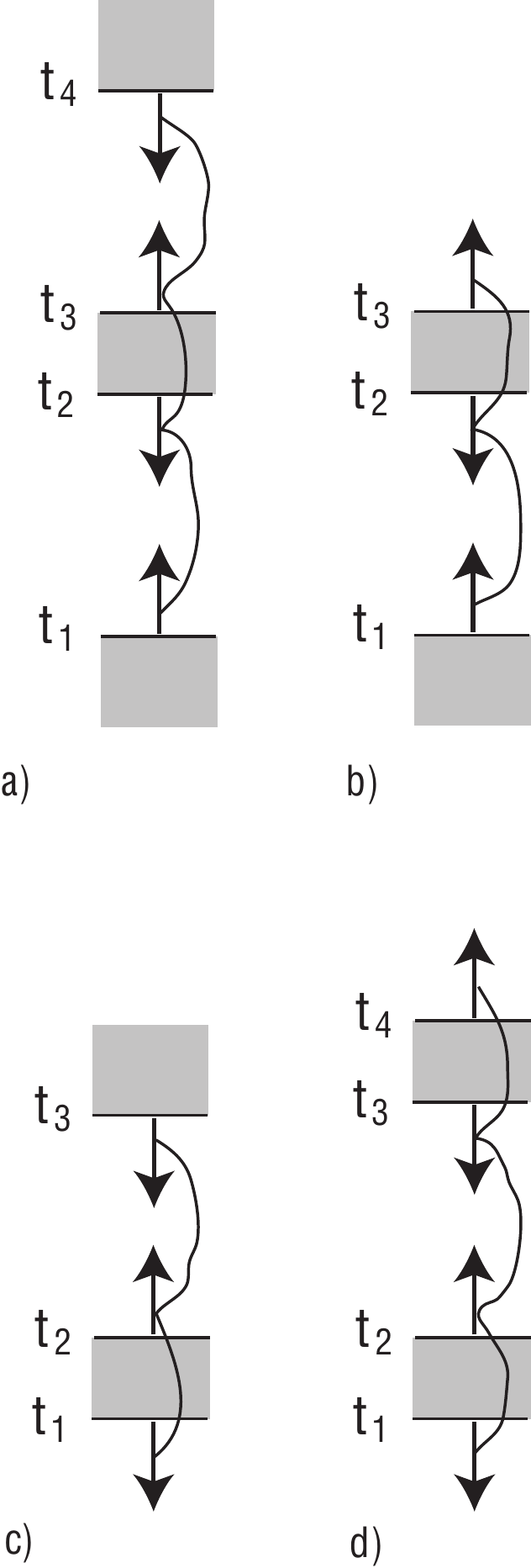} \caption {Different multi-time situations. In (a) both the past and the future
are well defined, i.e. they are part of "preparation" stages. In (b) the future is uncertain, in (c) the past is
uncertain while in (d) both the future and the past are uncertain.}
\end{figure}

We are now ready to state the basic result of our paper. Let ${\Psi}$ denote a state in ${\cal H}$.

\bigskip
\noindent {\bf Theorem: } In the case of multiple periods of preparation and measurements, any physical state of
a quantum system can be described by a vector $\Psi$ in ${\cal H}$
 or by mixtures of such vectors. Furthermore, to any vector or any  mixture of vectors in $\cal H$
 corresponds a physical state of the system.

\bigskip
\noindent In the above, the word "mixture" is taken to have the same two different meanings as in standard
quantum mechanics: (a) the preparer throws a die and prepares a different multi-time state for each outcome;
when the preparer gives us the state but doesn't inform us about the outcomes of the die, from our point of view
we have a mixture and (b) the multi-time state of the system is entangled with an ancilla.

\bigskip
\noindent It is  important to note that, unlike in standard quantum theory, we do not require the multi-time
states to be normalized. This is because there is no advantage in normalizing the multi-time states.  Indeed,
normalization of multi-time states does not automatically imply normalization of the probabilities of
measurement outcomes. Normalization of probabilities is an issue that can only be resolved when it is known what
measurements were actually performed. For any given set of measurements the state $\Psi$ prescribes only the
{\it relative} probabilities of the different outcomes of the measurements and the normalization of the
probabilities is then calculated so that the total probability is 1. Ultimately this stems from the fact that
the overall probability to prepare such a state depends on the probabilities of success of the different
post-selections involved, and these probabilities depend not only on what happens during the preparation times
but also on the measurements to which the state is subjected. This is different from the case of ordinary
one-time states which are prepared in advance; the probability of preparation is in this case equal to 1 and it
is independent of the measurement to which the system is thereafter subjected.

Finally, note that when discussing the case of multiple quantum systems, we may have a different number of
preparation and measurement stages for each system. For example \beq |\Psi_3\ra_{t_3}^A~{}_{t_2}^A\la
\Psi_2|~{}|\Psi_1\ra^A_{t_1}|\Phi\ra^B_{t_1}\eeq represents a state of two quantum systems, $A$ and $B$ in which
system $A$ is subjected to two preparation stages, from $t=-\infty$ to $t_1$ and from $t_2$ to $t_3$ while
system $B$ is subjected to a single preparation stage, from $t=-\infty$ to $t_1$. This idea generalizes easily
for multiple particles and multiple times.

\section{5. Measurement probabilities for multi-time states}

As in the case of two-time states, the meaning of the multi-time states is defined by the probabilities they
yield when the system is subjected to measurements. The probabilities for the outcomes of different measurements
are obtained from multi-time states in a very similar way to that in which they are obtained from one- and
two-time states. Consider first the case of detailed POVMs. To obtain the probability of a given outcome we
must:

\begin{itemize}
\item Step 1. Act on the multi-time state with the corresponding Krauss operators, i.e. insert the Krauss operators
in the appropriate slots and make all the scalar products with the  bra and ket vectors to which they apply.
(Note that if in a certain measurement period nothing is done, this corresponds to a Krauss operator that is
simply the identity)
\item Step 2. Compute the norm-squared of the resulting vector. Note that the four cases discussed in the previous section
(i.e. uncertain or
well-defined future and past) are slightly different: Indeed, after acting with the Krauss operators we end
either with a ket, a bra, a superposition of tensor products of a ket and a bra or just a complex number.
Computing the norm  has to be done in the appropriate way.
\item Step 3. Normalize the probabilities. That is, do steps 1 and 2 for each particular outcome - this will
determine the relative probabilities of the outcomes. To obtain the absolute probabilities divide all the
relative probabilities by their sum.
\end{itemize}

We now consider two examples. First consider the four-time state corresponding to the situation illustrated in
fig. 4a in which there is a well-defined past and future (determined by the initial preparation and final
post-selection) and two measurement periods ($t_1<t<t_2$) and ($t_3<t<t_4$). The multi-time state $\Psi$ for
this example is a vector in the Hilbert space ${\cal H}={\overrightarrow{\cal
H}}_{t_4}\otimes{\overleftarrow{\cal H}}_{t_3}\otimes{\overrightarrow{\cal H}}_{t_2}\otimes{\overleftarrow{\cal
H}}_{t_1}$ and can be expanded in terms of basis states as \beq\sum_{ijkl}\alpha_{ijkl}{~}_{t_4}\la
l|~|k\ra_{t_3}{}_{t_2}\la j|~|i\ra_{t_1}\label{four_times}\eeq

Let us denote the Krauss operators acting in the first and second measurement periods by $A_{\mu}$ and $B_{\nu}$
respectively, where $\mu$ and $\nu$ denote the corresponding results.

Acting on the state (\ref{four_times}) with the Krauss operators according to step 1 above, we obtain
\beq\sum_{ijkl}\alpha_{ijkl}{}_{t_4}\la l|B_{\nu}|k\ra_{t_3}{}_{t_2}\la j|A_{\mu}|i\ra_{t_1}\eeq which is a
complex number. According to step 2, the relative probability to obtain the results $\mu$ and $\nu$ is the
norm-squared of this complex number. Dividing these relative probabilities by their sum (step 3) we obtain the
absolute probabilities of the results $\mu$ and $\nu$, \beq{\rm Prob}(\mu, \nu)={1\over
N}\left|\sum_{ijkl}\alpha_{ijkl}{}_{t_4}\la l|B_{\nu}|k\ra_{t_3}{}_{t_2}\la
j|A_{\mu}|i\ra_{t_1}\right|^2\label{prob_case 1}\eeq where $N$ is such that $\sum_{\mu,\nu}{\rm Prob}(\mu,
\nu)=1$.

A second example corresponds to the situation illustrated in fig 4d.  In this example, both the future and the
past are uncertain, i.e. they belong to the experimentalist who performs measurements not to the preparer. There
are three measurement periods ($t<t_1$),  ($t_2<t<t_3$ ) and ($t_4<t$). The four-time state corresponding to
this situation is \beq\sum_{ijkl}\alpha_{ijkl}|l\ra_{t_4}{}_{t_3}\la k|~| j\ra_{t_2}{}_{t_1}\la i|.\eeq  Let
$A_{\mu}$, $B_{\nu}$ and $C_{\xi}$ denote the Krauss operators corresponding to the measurements performed in
the three measurement periods and $\mu$, $\nu$ and $\xi$ denote the corresponding results. Then the first step
is to act on the state with the Krauss operators. The result of acting with the Krauss operator and making all
the contractions (all the scalar products) is \beq\sum_{ijkl}\alpha_{ijkl}C_{\xi}|l\ra_{t_4}{}_{t_3}\la
k|B_{\nu}|j\ra_{t_2}{}_{t_1}\la i|A_{\mu}\eeq which is a tensor product between ket and bra vectors
corresponding to the initial and final time respectively. Indeed, note that in the above formula ${}_{t_3}\la
k|B_{\nu}|j\ra_{t_2}$ is just a complex number while $C_{\xi}|l\ra_{t_4}$ and ${}_{t_1}\la i|A_{\mu}$ are
un-contracted vectors. According to step 2, the relative probability to obtain the results $\mu$, $\nu$ and
$\xi$ is the norm-squared of this vector. Dividing these relative probabilities by their sum (step 3) we obtain
the absolute probabilities of the results \barr&&{\rm Prob}(\mu, \nu, \xi)={1\over
N}\left|\left|\sum_{ijkl}\alpha_{ijkl}C_{\xi}|l\ra_{t_4}{}_{t_3}\la k|B_{\nu}|j\ra_{t_2}{}_{t_1}\la
i|A_{\mu}\right|\right|^2\nonumber\\ &=&{1\over N}\sum_{ijkli'j'k'l'}\alpha^*_{i'j'k'l'}\alpha_{ijkl}~
{}_{t_4}\la l'|C^{\dagger}_{\xi}C_{\xi}|l\ra_{t_4} {}_{t_2}\la
j'|B^{\dagger}_{\nu}|k'\ra_{t_3}\times\nonumber\\&\times&{}_{t_3}\la k|B_{\nu}| j\ra_{t_2}{}_{t_1}\la
i|A_{\mu}A^{\dagger}_{\mu}|i'\ra_{t_1}\label{prob_case 2},\earr where $N$ is such that $\sum_{\mu,\nu,\xi}{\rm
Prob}(\mu, \nu, \xi)=1$.

\bigskip
It is important to note that when dealing with multi-time states that describe a situation with multiple
measurement periods, in order to be able to predict the probabilities for the outcomes of a given measurement we
need, in general, information about {\it all} the measurement periods, not only the ones when the measurement
takes place. Indeed, it is easy to see that what happens in other periods may influence the (relative)
probabilities of the different outcomes. Consider for example the four-time state $\sum_i
{}_{t_4}\la\Phi|~|i\ra_{t_3}{}_{t_2}\la i|~|\Psi\ra_{t_1}$ where the states $|i\ra_{t_3}$ and $_{t_2}\la i|$
form complete bases in their Hilbert spaces. This state describes a situation with two measurement stages, from
$t_1$ to $t_2$ and from $t_3$ to $t_4$. Suppose now that a measurement takes place from $t_3$ to $t_4$, and
suppose also that during the period from $t_1$ to $t_2$ some action is performed on the system, say a unitary
evolution $U$.  Then the probabilities ${\rm Prob}(k)$ turn out to be \beq {\rm Prob}(k)={1\over
N}|\la\Phi|A_kU\Psi\ra|^2\eeq where $A_k$ are the corresponding Krauss operators.  Clearly these probabilities
depend on $U$. Similarly, also in the case when the measurement takes place first, (i.e. measurement between
$t_1$ and $t_2$ and unitary evolution between $t_3$ and $t_4$, the probabilities are also influenced by $U$. In
this case \beq {\rm Prob}(k)={1\over N}|\la\Phi| UA_k\Psi\ra|^2.\eeq Basically, what happens during one time
period influences what happens during another time period via the correlations between the vectors associated
with these periods. The only case when we don't need to know information about all the periods is when some
periods effectively decouple from the rest, i.e, when the vectors that refer to these measuring periods are not
entangled with vectors from any other measurement period (see fig 5). In this case we can reduce the multi-time
state to an effective state covering only the connected periods of interest.

\begin{figure}[h] \includegraphics[scale=0.5]{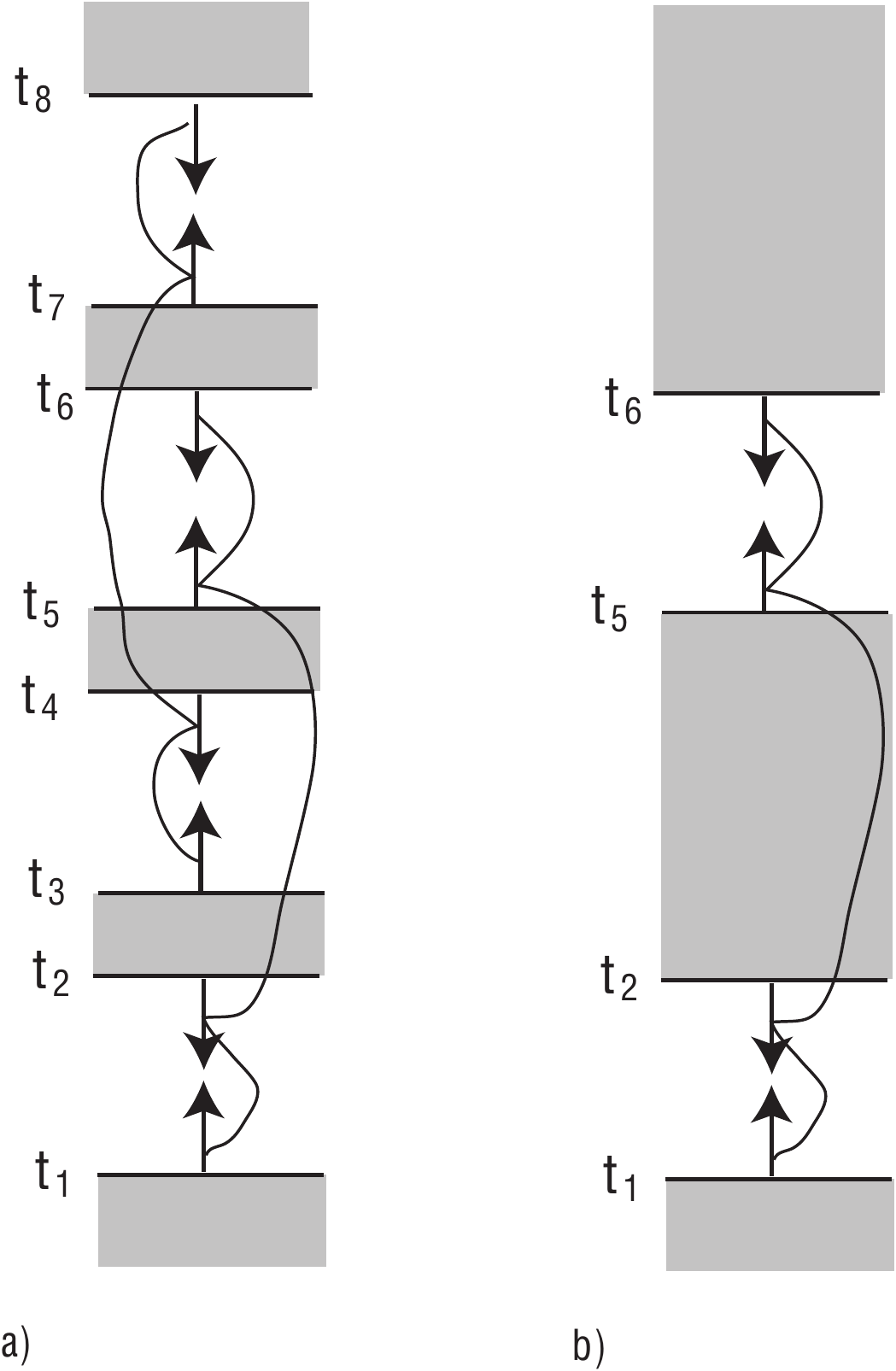} \caption {(a) The second and the fourth "measurement" stages,
i.e. from $t_3$ to $t_4$ and from $t_7$ to $t_8$ are correlated with each other but not with the other two
``measurement" stages. If we are interested in measurements that occurred during the first and the third stages,
the second and the fourth stages are irrelevant. (b) An effective state describing the periods $t_1$ to $t_2$
and $t_5$ to $t_6$ can be obtained simply by ignoring the other stages. } \end{figure}

Finally, the formalism can be made far more compact in the following way. When there are multiple measurement
periods, each characterized by its Krauss operator, we can define a global Krauss operator as the tensor product
of the individual operators corresponding to the different measurement periods. For example, when there are two
measurement periods, such as in the first example above, one described by $A_{\mu}$ and one by $B_{\nu}$ we can
define the total Krauss operator $K_{\lambda}=A_{\mu}\otimes B_{\nu}$ where the index $\lambda$ describes now
the outcome of the two measurements and is, in this case, nothing other than the pair ($\mu$,$\nu$). Then the
probability formula is

\beq {\rm Prob}(\lambda)={1\over N}\left|\left| K_{\lambda}\cdot
\Psi\right|\right|^2\label{krauss_probabilities}\eeq with $N$ such that $\sum_{\lambda}{\rm
Prob}(\lambda)=1\label{probability_formula1}$. Here by the dot product $K_{\lambda}\cdot \Psi$, we simply mean
that every bra (ket) vector belonging to $K_{\lambda}$ is contracted with the ket vector belonging to $\Psi$ and
corresponding to the same time and the contraction is the scalar product. This formula is the direct equivalent
of the well-known formula for determining the probability of a von Neuman measurement in a standard one-time
experiment, \beq {\rm Prob}(\lambda)={1\over N}\left|\left| P_{\lambda}|\Psi\ra\right|\right|^2\eeq where
$P_{\lambda}$ is the projector associated to the eigenvalue $\lambda$ of the measured observable and where
$N=\la\Psi|\Psi\ra$.

\bigskip
\noindent

\section{6. Preparing multi-time states. I}

There are many (infinite) ways of preparing multi-time states. Here we will present one particular method, which
is a generalization of the last method of preparing two-time states presented in section (3).

In this section we discuss multiple-time states in which the first time corresponds to a ket vector, that is, in
which the whole experiment starts with a preparation period. The cases that start with a measurement period are
discussed in the next section. We exemplify our method for an arbitrary  4-times state (fig 4a); generalizations
are obvious. The preparation procedure is illustrated in fig.6.

Consider the 4-time state \beq \sum_{ijkl}\alpha_{ijkl}~{}_{t_4}\la l|~|k\ra_{t_3}{}_{t_2}\la
j|~|i\ra_{t_1}\label{general_four_times}\eeq

We start by using three ancillas and preparing at $t_1$ the state

\beq \sum_{ijkl}\alpha_{ijkl}|i\ra_{S}|j\ra_{A1}|k\ra_{A2}|l\ra_{A3}\label{four_time_template}\eeq
which is a map of the desired state (\ref{general_four_times}). The ancillas are kept undisturbed
except when we use them to transfer their states onto the system. The transfer is performed via
post-selection of maximally entangled states and SWAP operations.

The role of maximally entangled states as channels for transforming ket states of the ancilla into bra vectors
of the system was discussed in section (3) and illustrated in fig 3. The swap operation has a similar role.
Indeed, the swap $S_{1,2}$ is a unitary operator that swaps the states of two quantum systems, S and A \beq
S_{SA}=\sum_{ij} |j\ra_S|i\ra_A{}_A\la j|{}_S\la i|.\eeq Note that this operator can also be written as \beq
S_{SA}=\bigl(\sum_{i} |i\ra_A{}_S\la i|\bigr)\bigl(\sum_{j}|j\ra_S{}_A\la j|\bigr)\eeq which is a product of two
mathematical objects, each of them looking like a maximally entangled state, but one in which a ket is entangled
with a bra. The swap operator then represents two entangled channels, one in which the forward-in-time
propagating state of the system is entangled with the backward-in-time propagating state of the ancilla and one
in which the forward-in-time propagating state of the ancilla is entangled with the backward-in-time propagating
state of the system. In particular the swap operator allows for the transfer of ket vectors of ancilla into ket
vectors of the system and of bra vectors of the ancilla into bra vectors of the system.

The overall procedure for preparing the state (\ref{general_four_times}) is the following:
\begin{itemize}
\item At $t_1$ prepare the entangled state (\ref{four_time_template}) of the system and of the ancillas.
\item At time $t'$, $t_2<t'<t_3$ perform the swap operation $S_{S,A2}$
between the system and ancilla A2. The system is kept undisturbed at all other times between $t_2$
and $t_3$.
\item At time $t''$, $t'<t''<t_4$ perform a Bell operator measurement on ancillas $A1$ and $A2$  and post-select the
maximally entangled state $|\Phi^+\ra_{A1,A2}$
\item At $t_4$ perform a Bell operator measurement on the system $S$ and ancilla $A3$  and post-select the
maximally entangled state $|\Phi^+\ra_{S,A3}$
\end{itemize}

The resulting state of the system and ancillas is \beq _{~~~t_4}^{S,A3}\la \Phi^+|~{}_{~~~~t''}^{A1,A2}\la
\Phi^+|~S_{t'}^{S,A2}~\sum_{ijkl}\alpha_{ijkl}|i\ra_{t_1}^{S}|j\ra_{t_1}^{A1}|k\ra_{t_1}^{A2}|l\ra_{t_1}^{A3}\eeq
By contracting the states of the ancillas (i.e. by making the appropriate scalar products) and by propagating in
time (without any change, since the system is undisturbed during these times) the state of the system, the bra
from $t'$ to $t_2$ and the ket from $t'$ to $t_3$  we obtain the desired state (\ref{general_four_times}). The
procedure is illustrated in fig 6. There the transfer of the ancilla states onto the system can be seen clearly.

\begin{figure}[h]
\includegraphics[scale=0.8]{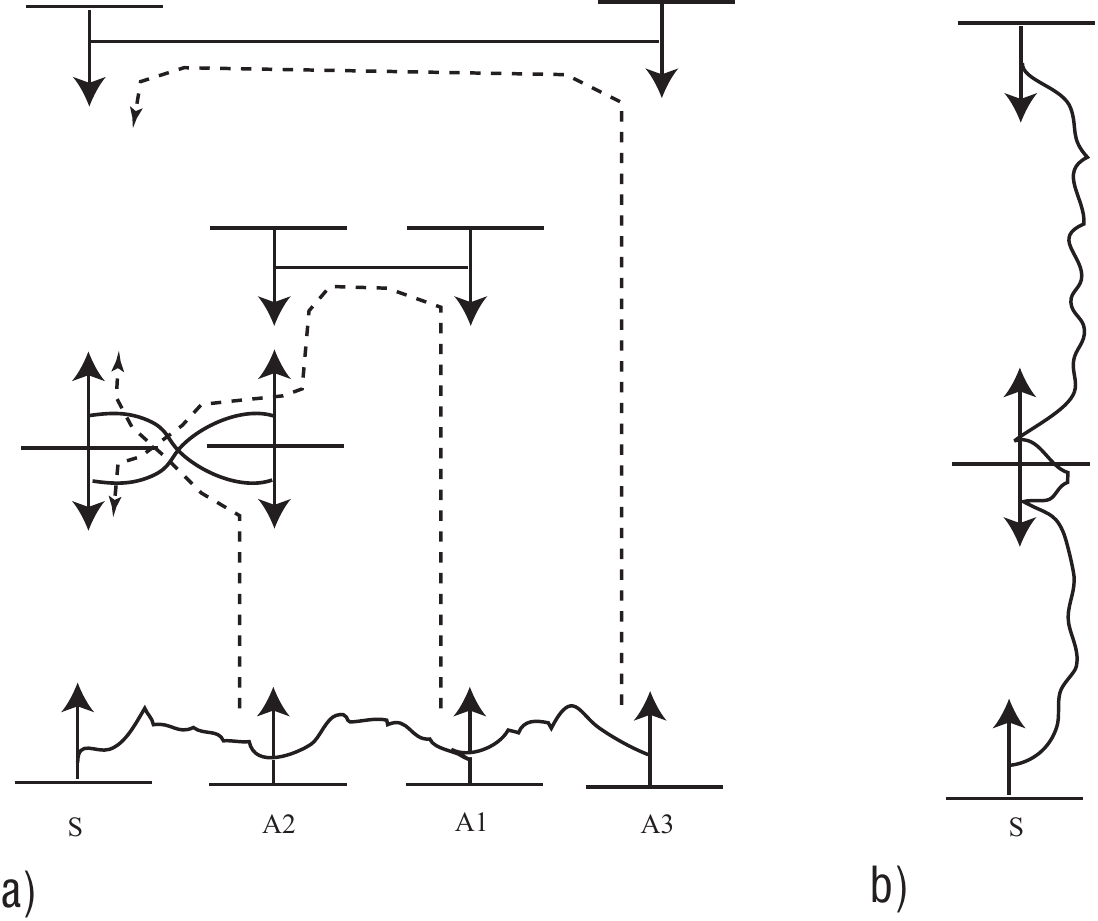} \caption {(a) The system ${\rm S}$ and the three ancillas, ${\rm A}_1$, ${\rm
A}_2$ and ${\rm A}_3$ start in an entangled state that gets transferred onto the system via the interactions of
the system and the ancillas and via appropriate post-selections. The continuous line describes maximal
entanglement while the wiggled line describes arbitrary entanglement and the arrows represent states propagating
``forward" and ``backward" in time, i.e. ket and bra vectors. Note the SWAP interaction between the system ${\rm
S}$ and the ancilla ${\rm A}_2$; it is nothing other that maximal entanglement between bra states of the system
and ket states of the ancilla and vice-versa. The dotted line illustrates how entanglement is transferred from
the ancillas onto the system. The diagram (b) illustrates the same situation as (a) but from the point of view
of the system alone.}
\end{figure}

Preparing a state for the case when the first period is a preparation and the last period  is a measurement
period, i.e. a state in which both the first and the last vectors  are  kets is done by a simple modification of
the procedure described above. Consider for example the 3-time state \beq
\sum_{ijk}\alpha_{ijk}~|k\ra_{t_3}{}_{t_2}\la j|~|i\ra_{t_1}\label{general_three_times}\eeq We prepare it in the
same way as the 4-time state above, only that the last ancilla, and therefore all the actions involving it, are
missing. That is we start from the state \beq
\sum_{ijk}\alpha_{ijk}|i\ra_{S}|j\ra_{A1}|k\ra_{A2}\label{three_time_template}\eeq and we perform the exact
procedure described above, except the final measurement at $t_4$.

\section{7. Past and future boundary conditions}

In the previous section we discussed experiments which start with a preparation stage. Correspondingly, the
multi-time states that describe them start with a ket vector. However, we mentioned in our general theorem that
we can also consider experiments that start with a measurement stage, and thus the corresponding multi-time
states start with a bra vector. At first sight this seems puzzling. Indeed, there is always some state prepared
in the remote past, either explicitly prepared by the experimentalist or naturally occurring. So it seems that
we should always start with a ket vector. The key however is to realize that this problem can be avoided if we
make the past ``neutral", i.e. if we arrange a situation such that all states coming from the remote past
towards our experiment are equally probable \cite{backward_evolving}. In other words, a neutral past is one in
which the initial state of the system is not any pure state but an equal mixture of all possible states, i.e.
(up to normalization) the identity density matrix. This can be done for example by actually starting the
experiment with a preparation stage in the remote past,  in which we maximally entangle the system with an
ancilla.

An example suffices -all multi-time states starting with a bra vector can be constructed in a similar way.
Consider the one-time state $_{t_1}\la\Psi|$ which is supposed to describe the situation illustrated in fig 7
where the experiment consists of a measurement period followed by a preparation.

\begin{figure}[h]
\includegraphics[scale=0.8]{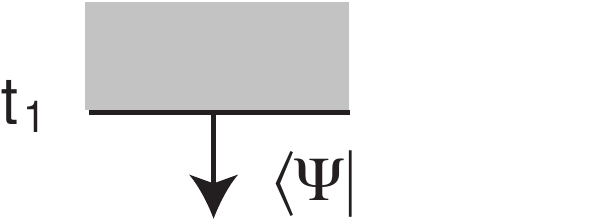} \caption {A simple one-time state with uncertain past.}
\end{figure}

According to our definitions, the meaning of this state is that if during the
measurement period we perform a detailed POVM described by the Krauss operators $A_k$, the
probability to obtain the outcome $k$ is given, up to normalization, by the norm
\cite{normalization} of the vector  $_{t_1}\la\Psi|A_k$, i.e. \beq {\rm Prob}(k)={1\over
N}\la\Psi|A_kA^{\dagger}_k|\Psi\ra.\eeq

A procedure for obtaining  $_{t_1}\la\Psi|$ is to prepare the pre and post selected state of the system and
ancilla \beq _{t_1}^S\la\Psi|~|\Phi^+\ra^{SA}_{t_0}\eeq where $t_0<t_1$ and where $|\Phi^+\ra^{SA}$ is the
maximally entangled state (\ref{maximally_entangled}) .  Note that the ancilla $A$ is then left unmeasured. One
can explicitly see that this state is equivalent to $_{t_1}\la\Psi|$. Indeed, \barr &&{\rm Prob}(k)={1\over
N}{}_{t_0}^{SA}\la\Phi^+|A_k^{\dagger}|\Psi\ra^S_{t_1}{}_{t_1}^S\la\Psi|A_k|\Phi^+\ra^{SA}_{t_0}=\nonumber\\&=&
{1\over N}\sum_n{}_{t_0}^{A}\la n|{}_{t_0}^{S}\la
n|A_k^{\dagger}|\Psi\ra^S_{t_1}{}_{t_1}^S\la\Psi|A_k\sum_m|m\ra^{S}_{t_0}|m\ra^{A}_{t_0}\nonumber\\&=&{1\over
N}\sum_{nm} \delta_{nm}~{}_{t_0}^{S}\la
n|A_k^{\dagger}|\Psi\ra^S_{t_1}{}_{t_1}^S\la\Psi|A_k|m\ra^{S}_{t_0}=\nonumber\\&=&{1\over
N}\sum_n{}_{t_1}^S\la\Psi|A_k^{\dagger}|n\ra_{t_0}^S{}_{t_0}^{S}\la
n|A_k^{\dagger}|\Psi\ra^S_{t_1}\nonumber\\&=&{1\over N}{}^S\la\Psi|A_kA^{\dagger}_k|\Psi\ra^S.\earr

Incidentally, this means that we can view the state $_{t_1}\la\Psi|$ both as a one-time pure state
and as a two-time mixture, (in which the kets at $t_0$ come with equal probability).

It is worth at this point looking in more detail at the ``future boundary condition" as well. By analogy with
the past boundary condition, we conclude that the future is akin to the post-selection of the identity density
matrix. In other words, we can view the standard one-time state $|\Psi\ra_{t_1}$ either as a pure one-time state
or as a two-time mixture (in which the bra vectors at $t_2>t_1$ come with equal probability).

To conclude the last two sections, we showed that any multi-time state can be prepared. There are many ways to
prepare
them, and the general method presented here may not be the most efficient, that is, the probability for
the success of all the required post-selections may not be optimal, Indeed, we did not make an optimality study
here. However, the main point, namely that all these states are possible, has been made.

\section{8. Particular examples of multi-time states}

An interesting case is the two-time state \beq \sum_{i=1}^n |i\ra_{t_2} ~_{t_1}\la i|\eeq where the vectors
$|i\ra_{t_2}$ and $_{t_1}\la i|$ respectively form complete orthonormal bases in  ${\overleftarrow{\cal H}}_2$
and ${\overrightarrow{\cal H}}_1$ respectively. Here the vectors propagating backward in time at $t_1$ are
completely correlated with those propagating forward in time at $t_2$, (i.e. the bra vectors at $t_1$ and the
kets at $t_2$) are ``maximally" entangled. In effect they form an identity operator. (Note that this is very
similar to the ordinary entanglement of two particles in a singlet type state, but here it is entanglement
between bra and ket vectors and represents total correlations in all possible basis while total correlations are
impossible in the case of entanglement between two sets of ket states - the singlet state represents total
anti-correlation not total correlation.) Most importantly, this state can be prepared by simply leaving the
system unperturbed between $t_1$ and $t_2$. In this case any information reaching $t_1$ is then propagated to
$t_2$. For example the state \beq \sum_{i=1}^n |i\ra_{t_2} ~_{t_1}\la i|~
|\Psi\ra_{t_0}\label{postselect_identity} \eeq is (up to normalization) nothing other than the standard state
$|\Psi\ra_{t_0}$ as one can see by verifying that all the probabilities for all the possible measurements are
the same for (\ref{postselect_identity}) and for $|\Psi\ra_{t_0}$. This example contains a most important
message: a time interval when nothing happens, such as between $t_1$ and $t_2$ here, is equivalent to a
preparation in which the backward-in-time and the forward-in-time propagating vectors emanating from this time
interval are ``maximally" entangled.

Another interesting state is \beq \sum_i {}_{t_2}\la i|~|i\ra_{t_1}\label{closed_time_loop}.\eeq Here the
vectors propagating forward in time at $t_1$ are completely correlated (i.e. maximally entangled) with those
propagating backwards in time at $t_2$. This state represents a ``closed time loop" - any information that
reaches time $t_2$ is "propagated" back to time $t_1$.

\section {9. Multiple-time measurements}

Up to this point when we discussed measurements we considered the usual quantum mechanical measurements, such as
measuring the observable $C$ at time $t$. But such measurements are very simple in the sense that they are
``one-time" measurements. One can consider far more complex measurements, namely multi-time measurements
\cite{multi-time-measurements}; it is very natural to consider such measurements here, when discussing
multi-time states.

A simple example of a two-time observable is $\sigma_x(t_1)-\sigma_x(t_2)$, the difference between the
x-component of the spin of a spin 1/2 particle at two different times. The important thing to note is that this
is an observable that gives the value zero in the case when the x-component of the spin is the same at the two
times, but doesn't offer any information about the actual value of the x-component. Measuring this operator is
therefore {\it not} equivalent to measuring the x-component of the spin at $t_1$ , followed by another
measurement at $t_2$ and finally subtracting the values of the results. Indeed, such a measurement would yield
too much information: it would tell the actual value of the spin at the both times, not only the difference. How
to measure such an observable has been described in \cite{multi-time-measurements} and we describe it here for
completeness.

Two ways to accomplish the above task are the following. In the first method we use a single
measuring device that we couple to the spin twice, once at $t_1$ and once at $t_2$. Following the
von Neumann measuring procedure \cite{von_Neumann} we consider a measuring device consisting of a
pointer whose position is denoted $q$ and its  conjugate momentum $p$.  The initial state of the
measuring device is the pointer indicating zero, i.e.  $|q=0\ra$. The measuring device interacts
with the spin via the interaction hamiltonian \beq H_{int}=\delta(t-t_1)p\sigma_x-
\delta(t-t_2)p\sigma_x.\eeq

The first time the coupling is such as to shift the pointer's position $q$ by an amount proportional to
$\sigma_x$ and the second time to shift it proportional to $-\sigma_x$.  Indeed, the time evolution
corresponding to the first interaction is $U(t_1)=e^{-ip\sigma_x}$ which is a shift operator shifting $q$ by the
value $\sigma_x$ while the evolution corresponding to the second interaction is a shift operator
$U(t_2)=e^{ip\sigma_x}$ representing a shift of $q$ by $-\sigma_x$. Assuming that during the time interval
between the two measurements the pointer is preserved in an undisturbed quantum state (i.e.  the effective
hamiltonian of the measuring device is zero between the two interactions with the spin) the Heisenberg equations
of motion show that \beq q_{final}=q_{initial} +\sigma_x(t_1)-\sigma_x(t_2).\eeq  As the initial position of the
pointer is known, $q_{initial}=0$, the final value of $q$ indicates $\sigma_x(t_1)-\sigma_x(t_2)$. Furthermore,
note that since we did not read the position of the pointer after the first interaction, when the whole
measurement is finished we no longer have the possibility of finding out what $\sigma_x(t_1)$ was. Also we
cannot find out what $\sigma_x(t_2)$ was because we don't know the position of the pointer before the second
interaction.

The second way to perform such a measurement involves two independent measuring devices one interacting with the
spin at $t_1$ and the other interacting at $t_2$. Let the two pointers be described by $q_1$,$p_1$ and
$q_2$,$p_2$ respectively and let the interaction Hamiltonian be \beq H_{int}=\delta(t-t_1)p_1\sigma_x-
\delta(t-t_2)p_2\sigma_x.\eeq To ensure that we do not get any information about the spin at $t_1$ and $t_2$ but
only about the difference we prepare the pointers in the entangled state $|q_1-q_2=0, p_1+p_2=0\ra$. In this
state the initial position of each pointer is completely uncertain so by reading their indications after the
measurement we cannot infer $\sigma_x(t_1)$ and $\sigma_x(t_2)$ separately, only their difference.

Now, although  in the discussion above we described in detail how to measure
$\sigma_x(t_1)-\sigma_x(t_2)$ it is  important to note that in order to predict the probabilities
for the different outcomes  we do not need to know the specific way in which the measurement is
implemented; just knowing the state and the observable itself is enough. This is similar to the
case of ordinary one-time variables usually studied in quantum mechanics.  For example when
considering  an ideal measurement of an observable $C$ we don't need to describe the entire
measuring procedure. We just use the state which is measured and the projectors on the different
eigenvalues of $C$.

For example, the probabilities of the different outcomes of an ideal measurement of
$\sigma_x(t_1)-\sigma_x(t_2)$ are obtained by using the projectors corresponding to its  different eigenvalues.
The observable $\sigma_x(t_1)-\sigma_x(t_2)$ has three eigenvalues, +2, 0 and -2. The value +2 is obtained when
$\sigma_x$ is  ``up" at $t_1$ and ``down" at $t_2$. The corresponding projector  is \beq P_2=
|\dn\ra_{t_2}|\up\ra_{t_1}~_{t_2}\la \dn|_{t_1}\la \up|\label{proj_spin_diff 2}\eeq The projector corresponding
to -2 is \beq P_{-2}=|\up\ra_{t_2}|\dn\ra_{t_1}~_{t_2}\la \up|_{t_1}\la \dn|\label{proj_spin_diff_minus2}\eeq
Finally, the projector corresponding to 0 is \beq P_0= |\up\ra_{t_2}|\up\ra_{t_1}~_{t_2}\la \up|_{t_1}\la
\up|+|\dn\ra_{t_2}|\dn\ra_{t_1}~_{t_2}\la \dn|_{t_1}\la \dn|\label{proj_spin_diff 0}\eeq

The way to use these projectors is identical to the way the projectors for one-time measurements are used: we
insert them into the state, in the corresponding slots and make the scalar products. Then the probability to
obtain, say $\sigma_x(t_2)-\sigma_x(t_5)=0$ when the spin is, say, in the four-time state $ {_{t_6}\la \Psi|
}~{|\Phi\ra_{t_4}} {_{t_3}\la\Xi|}~{|\Theta\ra_{t_1}}$  with $t_1<t_2<t_3<t_4<t_5<t_6$ is

\barr&& {\rm Prob}(\sigma_x(t_2)-\sigma_x(t_5)=0)=\nonumber \\&{1\over N}&\big| {_{t_6}\la \Psi| }
|\up\ra_{t_5}{_{t_5}\la \up|}{|\Phi\ra_{t_4}} {_{t_3}\la\Xi|}|\up\ra_{t_2}{_{t_2}\la \up|}{\Theta\ra_{t_1}}
+\nonumber\\&+&{_{t_6}\la \Psi| } |\dn\ra_{t_5}{_{t_5}\la \dn|}{|\Phi\ra_{t_4}}
{_{t_3}\la\Xi|}|\dn\ra_{t_2}{_{t_2}\la \dn|}{\Theta\ra_{t_1}}\big|^2 \label{multi_time_measurements_prob}\earr
In the above formula we considered that the hamiltonian affecting the spin is zero; if this is not so we need to
add the corresponding unitary transformations.

\section{10. Preparing multi-time states. II}

In section(6) we presented a particular method (based on SWAPs and postselection of maximally entangled states)
that allows the preparation of any arbitrary multi-time state. It is important to note however that any
measurement can be used to prepare multi-time states. This is similar to the situation in the standard
discussions of quantum measurements, but the multi-time approach introduces a very important twist.

The usual case is the following. Suppose that the state of a system at time $t_1$ is $|\Psi\ra$ and then a
measurement is performed between $t_1$ and $t_2$. When the measurement is a detailed POVM and the  outcome $k$
is observed, the state of the system at $t_2$ becomes (up to normalization)  $|\Phi\ra=A_k|\Psi\ra$, where $A_k$
is the corresponding Krauss operator.

In the usual way of looking at preparations as described above, the role of the operator  $A_k$,  is to
transform the initial state into the final state. However, as we will now show, this way of looking at the
problem obscures the true role of $A_k$. The operator is not there in order to evolve the state, but it is {\it
part of the state itself}. A few examples will make this situation clear.

Suppose a quantum system was prepared at time $t_0$ in the state $|\Psi\ra_{t_0}$. Furthermore, suppose that
between times   $t_1$ and $t_2$ a measurement was performed and the outcome $k$ (corresponding to $A_k$) was
obtained. The result is the three-time state

\beq  A_k^{t_2, t_1}~ |\Psi\ra_{t_0}\label{krauss_plus_state}\eeq where we added upper indexes to the Krauss
operator to denote the times between which it acts.  To better understand the meaning of the above state, note
that any Krauss operator acting between  $t_1$ and $t_2$ can be written as \beq A_k^{t_2, t_1}=\sum_{i,j}
\alpha_{i,j} |j\ra_{t_2}~{}_{t_1}\la i|.\label{krauss_state}\eeq Indeed, any linear operator acting on ket
vectors at $t_1$ and transforming them into ket vectors at $t_2$ can be written in this form. Hence, explicitly
written, the state (\ref{krauss_plus_state}) is \beq \sum_{i,j} \alpha_{i,j} |j\ra_{t_2}~{}_{t_1}\la
i|~|\Psi\ra_{t_0}\eeq

In the above we considered the POVM performed on a quantum system that was prepared at $t_0$ in state
$|\Psi\ra_{t_0}$. The effect of adding information from the POVM was to expand the state from a one-time state
to a three-times state by simply adding the Krauss operator into the state. This procedure is however far more
general: Whatever a multi-time state is, if we are further told that a POVM was performed, we simply add the
corresponding Kraus operator into the description of the state (and therefore expand an $n$-time state into an
$(n+2)$-time state).

The true force of the formalism however only becomes clear when we consider multi-time measurements such as
those described in section 9. A multi-time measurement has no simple description in the standard quantum
formalism. There is no ordinary Krauss operator that simply propagates an initial state into a final state,
since there is no well defined ``final" state. Indeed, the measurement takes place at many times, and there can
be any other interactions in between. In the multi-time formalism however any multi-time measurement can be
described by Krauss operators - they are however multi-time operators. An example is the spin measurement
described in section 9. The Kraus operators corresponding to this measurement are the multi-time projectors
(\ref{proj_spin_diff 2} - \ref{proj_spin_diff 0}).  To obtain the state of the system given the outcome $k$ of
the POVM all we do is, again, just to insert the multi-time Krauss operator into the original multi-time state.
Fig 8 illustrates the procedure.

\begin{figure}[h]
\includegraphics[scale=0.5]{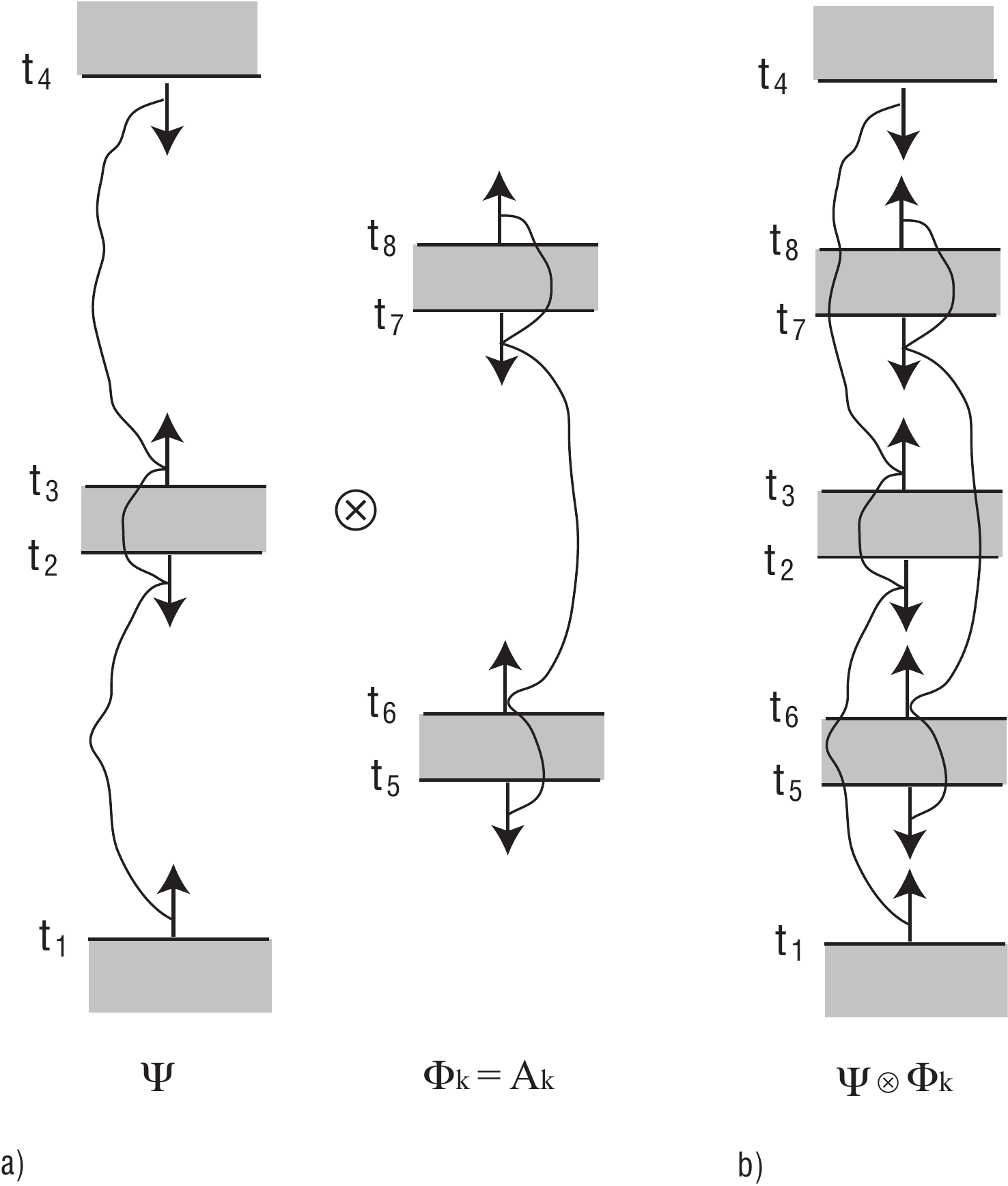} \caption { Using measurements for preparation. To the original state $\Psi$ we
add the information that the result of a POVM performed between $t_5$ to $t_6$ and $t_7$ to $_8$ yielded the
outcome $k$ corresponding to the Krauss operator $A_k$, The operator $A_k$ can also be viewed as a multi-time
state $\Phi_k$.  The new state that takes into account all the information is $\Psi\otimes\Phi_k$ which is
simply the composition of $\Psi$ and $\Phi_k$, as illustrated in (b)}
\end{figure}

\section {11. Operators versus states}

One of the main advantages of the multi-time formalism presented in this paper is to put states and operators on
an equal footing. Indeed, to start with, operators and multi-time states look formally identical - they are both
just superposition of tensor products of bra and ket vectors at different times. But this similarity is by no
means only superficial or coincidental. In the standard quantum mechanical formalism states are meant to
describe how the system was prepared while operators are meant to describe measurements performed on the system.
But physically preparations and measurements both involve exactly the same processes - interactions of the
system of interest with other quantum systems and/or with measuring devices. The multi-state formalism succeeds
in making this explicit.

As we argued in the introduction, the projector operators describing a von Neumann measurement (or indeed, more
generally, the Krauss operators) can be viewed as ``measurement states", in the sense that they encode all the
relevant information about the measurement. But we find it now very useful to think of both the ordinary
multi-time states  (that describe the way in which the system was prepared) and the measurement states (that
describe the measurements) on equal footing, as "histories".  This view allows a lot of flexibility.

Let  the state of the system be $\Psi$ and let us denote the Krauss operators $A_k$ that describe a given POVM
by $\Phi_k$ to emphasize that each of them can be interpreted as a state. Now, if we use the measurement as part
of the preparation, i.e. if in addition to the information that the system was prepared in the state $\Psi$ we
also are informed that we obtained the result $k$, then the new state of the system is simply the tensor product

\beq \Psi\otimes\Phi_k.\eeq where by the tensor product we  mean combining the two states, as described in the
previous section. What this formula tells us is that the total history is simply the combination of the two
histories.

On the other hand, suppose that we want to use a POVM not to prepare a state but to test it. That is, suppose we
ask, given the state $\Psi$ what are the probabilities to obtain different outcomes $k$? In general, of course,
there is no definite answer - the answer may depend on other things that may occur to the system meanwhile. For
example, suppose we are given the two-time state  $_{t_2}\la \Phi| |\Psi\ra_{t_1}$ and the POVM takes place
between two intermediate times, $t'$ and $t''$, $t_1<t'<t''<t_2$. Then,  the probabilities of the outcomes of
the POVM depend also on what happens between $t_1$ and $t'$ and between $t''$ and $t_2$ and therefore we cannot
determine them unless we are given this supplementary information.   But if we are given the whole information,
that is, if in effect the POVM covers the whole measurement period from $t_1$ to $t_2$ than we can predict its
results. In the operator language, as described in (\ref{krauss_probabilities}) we have to apply the different
Krauss operators to the state and compute the norms of the resulting vectors. On the other hand, we can
interpret the same formula as telling that  the probability is given, (up to overall normalization), by the norm
square of the scalar product between the two histories, the state of the system and the measurement state,
\beq{\rm Prob}(k)={1\over N}|\Phi_k\cdot \Psi|^2.\label{scalar_product}\eeq

This formula generalizes for arbitrary multi-time states and measurements. Of course, in order for the
probabilities to be well defined, the POVM must entirely cover all the measurement periods (or only some of the
measurement periods, in case they are disconnected from the rest- see the discussion at the end of section 5).
In case the POVM covers  all the measurement periods, then we use in the probability formula
(\ref{scalar_product}) the full state $\Psi$; otherwise we use the reduced state.

Finally note that depending on the past and future boundary conditions, the ``scalar product" of the two
histories is not always just a complex number but may also be a bra vector, a ket,  or a superposition of bra
and ket pairs (see the discussion in section 7). In those cases the ``norm square" of the scalar product is to
be taken as the norm of the resulting vectors.

In any case, conceptually, what the formula (\ref{scalar_product}) does is to generalize the usual notion that
when a system is in a state $|\Psi\ra$, the probability of finding it in the state $|\Phi\ra$ is the norm square
of the scalar product between $|\Psi\ra$ and the measured state $|\Phi\ra$, i.e.
$\left|\la\Phi|\Psi\ra\right|^2$.

\section{12. Measurements - open questions}

As far as the states of the system are concerned, the situation is completely solved: any superposition of
products of bra and ket vectors is a legitimate state of the system. Coming now to measurements, there are open
questions.

As discussed in previous sections, a measurement can be described in two different ways. One way, is to say
exactly how the measurement is performed. Of course, every measurement for which we are given the explicit
recipe of how to implement can, in principle, be performed. The second way of describing measurements is via its
Krauss operators. It is in connection with this latter way of describing measurements that there are very
interesting open problems.

In the case of ordinary one-time measurements, any set of Krauss operators (provided they fulfill the
normalization condition (\ref{povm_normalization}) represents a possible measurement. This is not the case for
multi-time measurements.  In fact there are two questions here.

Firstly,  is it the case that any superposition of products of bra and ket vectors as discussed above represents
a possible Krauss operator? In other words, given a Krauss operator, can we always find some multi-time
measurement such that this operator represents a particular outcome of the measurement? Or, to put it in a yet
other way, can every arbitrarily given history be implemented by a measurement?

Secondly,  a measurement is described not by a single history (i.e. a single Krauss operator) but by a whole set
of them. For example, an ideal von Neumann measurement is characterized by a complete basis of orthogonal
projectors. Then what are the conditions that a set of  histories must satisfy in order to describe a
measurement? That is, even if each Krauss operator in a set is legitimate, i.e. if each Krauss operator
separately it describes an outcome of a possible measurement, does the set of them describe a possible
measurement?

One major issue here is that measurements must obey causality.  That is, by acting on the system in the future
we shouldn't be able to change the probabilities of outcomes of measurements in the past. While this condition
is obeyed automatically for measurements that are sequences of one-time measurements, it is not the case that
any arbitrary set of legitimate histories obeys this constraint. But it is also possible that there are cases of
sets of Krauss operators that do not lead to causality violations but still there is no actual way to implement
them in quantum mechanics.

A somewhat similar situation is encountered when dealing with instantaneous non-local measurements. In that case
there are sets of legitimate Krauss operators that fulfill the normalization constraint but are not measurable
because they would lead to superluminal signaling \cite{popescu_vaidman}. There are also known cases
\cite{horodecki} when a set of Krauss operators is unmeasurable although it wouldn't lead to superluminal
communication but would allow for establishing non-local correlations stronger than allowed by quantum mechanics
(Popescu-Rohrlich type correlations \cite{popescu_rohrlich}). Finally there may other cases of non-measurable
sets of Krauss operators in which the reason for unmeasurability is different from the above. Coming back to
multi-time measurements, we expect to find similar behavior.

Partial answers to the above questions and other related problems are discussed in  \cite{tony}.

\section{13. Discussion: The flow of time}

So far in this paper we approached the idea of multiple-time states from a rather formal point of view and
avoided questions of interpretation. That is, we considered physical situations in which a quantum system is
subjected to multiple stages of preparation and measurement. We then asked, given the preparation, what is the
set of parameters that are relevant for inferring as well as possible the results of the measurements. What we
found is that these parameters can be expressed as vectors in a "multi-time" Hilbert space (which is the tensor
product of Hilbert spaces associated with each time boundary between preparation and measurement stages). Each
vector, or mixture of vectors, describes a possible physical situation, and each possible physical situation can
be described in this way. Clearly this is a basic fact about the structure of quantum mechanics and it is here
to stay, no matter what philosophical interpretation we may associate with these states. It is very tempting,
however, to go further and ask what does this all mean.

As we mentioned  in the introduction, trying to give a philosophical interpretation for multi-time states is
certainly not easy. Indeed, even the interpretation of the ordinary (one-time) quantum state is highly
controversial. We ourselves do not have one preferred interpretation of multi-time states - in fact we have two
of them, and we find both these points of view useful. We will describe here one of these points of view while
the other one, the ``block-time universe", is presented in a forthcoming paper \cite{heraclit}.

It is quite usual when thinking about the ordinary quantum state, to regard it not just as a static collection
of parameters associated to some preparation stage,  but to think that at each moment in time the system is
described by a ``state", i.e. by a ket vector, and that this state evolves in time, being affected by all the
interactions the system has. On one hand, one can view this ``evolution" as a simple mathematical procedure by
which we transform the parameters given at the preparation time $t_0$ into a more convenient form for computing
what happens at the moment of interest $t$. In effect, we simply interpret part of the measurement stage, namely
the period from $t_0$ to $t$ as being part of the preparation stage.  On the other hand,  one may  view the
state as a physical object that evolves in time, undergoes  collapses, etc. Obviously, although the
probabilities we compute using these two different notions of state are the same, there is a great conceptual
difference here - the state being a simple mathematical recipe for computing probabilities versus the state
having an objective physical existence.

But consider now the simple example illustrated in fig 2b. As far as the preparation is concerned, the system is
described by the two-time state $_{t_2}\la \Phi|~|\Psi\ra_{t_1}$. Suppose further that the moment of interest is
some time $t$, $t_1<t<t_2$. We can then mathematically ``evolve"  the vectors $|\Psi\ra$ forward and $\la\Phi|$
backward until they reach that moment, $t$. The (ket) vector $|\Psi\ra$  originates at $t_1$, it is determined
by the time boundary condition in the past, and ``evolves" toward the future. The  (bra) vector $\la\Phi|$
originates at $t_2$, it is determined by the time boundary condition in the future and ``evolves" toward the
past. Again, in effect all we do is to include the period from $t_1$ to $t$ and the period from $t$ to $t_2$
into the preparation stage instead of in the global measurement stage. On the other hand, we could think of the
vectors $|\Psi\ra$ and $\la\Phi|$ as having objective physical meaning. This view however implies a dramatic
conceptual change, far greater that that related to the interpretation of the standard quantum state.  Indeed,
the issue now is no longer only whether or not the quantum state has objective meaning or is just a mathematical
tool for computing probabilities. The issue is now that of the {\it flow of time}.

To start with, it is a quite trivial fact that if we acquire new information we can affect the probabilities of
events that happened in the past. This happens not only in quantum mechanics but in ordinary classical
probabilities as well. For example suppose we have a bag with an equal number of white and black balls and
extract one ball at random and put it, without looking, into a bag containing only black balls. The probability
that the ball is white is 1/2. But suppose we than extract a ball from the second bag and see that the ball is
white. In the light of this new information we can now infer that in this situation the probability that a white
ball was extracted from the first bag is actually 1 and not 1/2. The future information affects our knowledge
about the past, but there is nothing surprising about this. Similarly, there is nothing surprising about the
fact that post-selection at $t_2$ affects the probabilities for events that happened at the earlier time $t$.
So, as long as we view the vector $\la\Phi|$ just as a mathematical tool for calculating probabilities, it is
nothing surprising that it ``evolves" backward in time. But if $\la\Phi|$ has objective meaning, than we have to
admit that it really propagates backwards in time.

At first sight it appears that the idea of a state propagating backwards in time is ridiculous and should be
immediately abandoned. The example of the classical post-selection described above seems to show that an attempt
to interpret the change in the statistics of results of experiments due to post-selection as a true
backward-in-time influence is trivially wrong. However, we do feel that the situation is far more interesting in
quantum mechanics. Indeed, there is a fundamental difference between post-selection in the classical and quantum
cases. In the classical case, probabilities are only due to our {\it subjective} lack of knowledge. In
principle, we could have had {\it complete} information about the system from the initial moment, and then there
is no issue of probabilities and a future measurement doesn't really provide new information. On the other hand,
it is one of the most important aspects of quantum mechanics - perhaps {\it the} most important aspect - that
even when we have whole information about the past (say, we know the state $|\Psi\ra$ at $t_1$), in general we
still cannot predict with certainty the result of a later measurement. The later measurement  does therefore
yield truly new information about the system.  In other words, the future is not completely determined by the
past. Hence the whole notion of past and future in quantum mechanics is fundamentally different than in
classical mechanics, and the whole idea of time flow may need to be reconsidered.

Of course, as we emphasized from the very beginning of this paper, all our results are fully consistent with
ordinary quantum mechanics. In particular they could all be obtained using the traditional view of a single
quantum state evolving in time. But  we personally found it very useful to think of states propagating forward
and backward in time. In particular, during each measurement period we think of two vectors, a ket propagating
from the past time-boundary condition towards the future and a bra  propagating from the future time-boundary
condition towards the past. Each moment of time is therefore described by these two vectors \cite{two-vectors}.
Of course, more generally each time moment can be described by entangled bra and ket vectors or mixtures of
them.

Thinking of vectors propagating forward and backward in time opens many new possibilities that we found very
intriguing. In particular, one can ask about the possibility of having such time flow consistent with free-will.
As we show elsewhere, that it is consistent \cite{free-will}.
It is also possible to take forward and backward
in time propagation as a starting point for possible modifications of quantum mechanics. Finally, it is tempting
to try and apply the idea of multi-time states in cosmological context, in particular to speculate about the
possibility that the Universe has both an initial and a final state which are given independently of each other.

\section{Acknowledgments}

YA and LV acknowledges support by grant 990/06 of the Israel Science Foundation. LV acknowledges support in part
by the European Commission under the Integrated Project Qubit Applications (QAP) funded by the IST directorate
as Contract Number 015848 and . SP acknowledges support from the UK EPSRC grant GR/527405/01 and the UK EPSRC
QIP IRC project. JT acknowledges support by grant N00173-07-1-G008 of the Naval Research Laboratory.


\begin{thebibliography}{99}
\bibitem{normalization} Incidentally, note that the normalization factor in this case is {\it not} equal to 1.
Indeed, $N=\sum_k\la\Psi|A_kA^{\dagger}_k|\Psi\ra=\la\Psi|\sum_kA_kA^{\dagger}_k|\Psi\ra$ but in general
$\sum_kA_kA^{\dagger}_k\neq 1$ - it is only the expression $\sum_kA^{\dagger}_kA_k$ which is equal to 1 due ty
the normalization of POVMs., not the expression $\sum_kA_kA^{\dagger}_k$.
\end{thebibliography}
\end{document}